\documentclass[aps,prb,twocolumn,superscriptaddress,floatfix]{revtex4-2}
\usepackage{graphicx}
\usepackage{multirow}
\usepackage{color}
\usepackage{amsmath}
\usepackage{amsfonts}
\usepackage{amssymb}
\usepackage{graphicx}
\graphicspath{{Figures/}}
\usepackage{epstopdf}
\usepackage{empheq}
\usepackage{float}
\usepackage{cases} 
\usepackage{mathtools}
\usepackage{dcolumn}
\usepackage{bbold}
\usepackage{xfrac}
\usepackage{bm}
\usepackage{siunitx}
\usepackage{nicefrac}

\usepackage[dvipsnames]{xcolor}
\usepackage[colorlinks]{hyperref}
\hypersetup{
colorlinks=True,
linkbordercolor=White,
linkcolor=BrickRed,
citecolor=Blue,	 
}

\newcommand{\lsesTthree}{$\beta_\parallel($T$3)\,$}
\newcommand{\lsesTfour}{$\beta_\parallel($T$4)\;$}

\begin{document}

\title{Optimal operation of hole spin qubits}

\author{M. Bassi}
\affiliation{Univ. Grenoble Alpes, CEA, Grenoble INP, IRIG-Pheliqs, 38000 Grenoble, France.}
\author{E. A. Rodr\'iguez-Mena}
\affiliation{Univ. Grenoble Alpes, CEA, IRIG-MEM-L\_Sim, 38000 Grenoble, France.}
\author{B. Brun}
\author{S. Zihlmann}
\author{T. Nguyen}
\author{V. Champain}
\affiliation{Univ. Grenoble Alpes, CEA, Grenoble INP, IRIG-Pheliqs, 38000 Grenoble, France.}
\author{J. C. Abadillo-Uriel}
\affiliation{Univ. Grenoble Alpes, CEA, IRIG-MEM-L\_Sim, 38000 Grenoble, France.}
\affiliation{Instituto de Ciencia de Materiales de Madrid, CSIC, 28049 Madrid, Spain}
\author{B. Bertrand}
\author{H. Niebojewski}
\affiliation{Univ. Grenoble Alpes, CEA, LETI, Minatec Campus, 38000 Grenoble, France.}
\author{R. Maurand}
\affiliation{Univ. Grenoble Alpes, CEA, Grenoble INP, IRIG-Pheliqs, 38000 Grenoble, France.}
\author{Y.-M. Niquet}
\affiliation{Univ. Grenoble Alpes, CEA, IRIG-MEM-L\_Sim, 38000 Grenoble, France.}
\author{X. Jehl}
\author{S. De Franceschi}
\author{V. Schmitt}
\affiliation{Univ. Grenoble Alpes, CEA, Grenoble INP, IRIG-Pheliqs, 38000 Grenoble, France.}

\email{}

\date{\today}

\begin{abstract}

Hole spins in silicon or germanium quantum dots have emerged as a compelling solid-state platform for scalable quantum processors. Besides relying on well-established manufacturing technologies, hole-spin qubits feature fast, electric-field-mediated control stemming from their intrinsically large spin-orbit coupling \cite{Fang_2023, Burkard_2023}. This key feature is accompanied by an undesirable susceptibility to charge noise, which usually limits qubit coherence. Here, by varying the magnetic-field orientation, we experimentally establish the existence of ``sweetlines'' in the polar-azimuthal manifold where the qubit is insensitive to charge noise.  In agreement with recent predictions \cite{michal_2023}, we find that the observed sweetlines host the points of maximal driving efficiency, where we achieve fast Rabi oscillations with quality factors as high as 1200. Furthermore, we demonstrate that moderate adjustments in gate voltages can significantly shift the sweetlines. This tunability allows multiple qubits to be simultaneously made insensitive to electrical noise, paving the way for scalable qubit architectures that fully leverage all-electrical spin control. The conclusions of this experimental study, performed on a silicon metal-oxide-semiconductor device, are expected to apply to other implementations of hole spin qubits.
 
\end{abstract}

\maketitle

Since the first proof-of-concept demonstrations in silicon \cite{Maurand2016} and germanium \cite{Watzinger_2018} quantum dots, hole spin qubits have made significant strides in both operational performance and scalability \cite{Fang_2023}. Notable achievements include high-speed single\cite{Froning_NNano_2021, Wang_2022}- and two-qubit gates \cite{Hendrickx_Nature_2020, Geyer_2024}, singlet-triplet qubits \cite{Jirovec_2021, Liles_2024}, quantum dot arrays hosting up to ten qubits \cite{Wang_2024}, and the creation of four-qubit entangled states \cite{Hendrickx_Nature_2021}. Central to these advances is spin-orbit coupling, which enables fast electric control of spins but also increases sensitivity to electrical noise, limiting coherence times. Theoretical predictions and supporting experimental observations have identified special magnetic-field orientations, known as ``sweetspots'', where hole spins become first-order immune to electrical noise \cite{Piot_2022}. Operating hole spin qubits at these sweetspots significantly enhances Hahn echo coherence times, beyond several tens of microseconds in both natural silicon and germanium quantum dots \cite{Piot_2022, Hendrickx_2024}. While these advancements mark a clear breakthrough in improving qubit coherence, they also raise important questions about the characteristics of sweetspot operation and its feasibility in large-scale qubit architectures. 

How does the sweetspot condition manifests across the full range of polar and azimuthal magnetic-field angles? To what extent can noise resilience coexist with the ability to perform fast spin control? 
Recent theoretical predictions even suggest that optimal coherence and control speed can coexist \cite{Michal_PRB_2021, Mauro_2023}, with initial experimental evidence supporting this intriguing possibility \cite{carballido_2024}. Finally, considering the inherent variability in quantum dot properties, it is crucial to determine whether noise-resilient operation could be realistically established across large qubit arrays — a key requirement for the development of scalable quantum processors based on hole spin qubits. The above mentioned questions are addressed in this experimental work.  

In a static magnetic field $\bm{B}$, the up and down spin states split by a Zeeman energy $E_Z = \mu_B g^* |\bm{B}|$, where $\mu_B$ is Bohr’s magneton and $g^*$ is the effective gyromagnetic factor (in short, $g$-factor). Due to spin-orbit coupling, $g^*$ becomes dependent on the magnetic field orientation and on the local electric fields \cite{Crippa_PRL_2018}. Consequently, spin dynamics is governed by time-dependent electric fields, induced either by driving signals applied to the gates, or by charge noise. In general, the action of a gate-voltage modulation $\delta V$ is twofold (for a more detailed and rigorous discussion, see Methods and Supplementary Information). 
First, it can shift the Larmor precession frequency of the spin $f_L = E_Z/h$ (with $h$ the Planck constant). This so-called longitudinal response is to first-order proportional to the longitudinal spin-electric susceptibility (LSES) $\beta_\parallel= \partial f_L/\partial V$. When longitudinal fluctuations  originate from gate-voltage noise or, equivalently, charge fluctuations in the environment, they lead to dephasing due to random variations in the Larmor frequency.  Second, $\delta V$ can mix the up and down spin components leading to population changes. If the voltage modulation frequency is in resonance with the Larmor frequency, this so-called transverse coupling gives rise to coherent Rabi oscillations at frequency $f_R = \beta_\perp V_\text{ac}$, where $\beta_\perp$ is the transverse spin-electric susceptibility (TSES) and $V_{ac}$ the amplitude of the resonant drive.
It is, therefore, highly desirable to find operational sweetspots where the longitudinal coupling $\beta_\parallel$ is minimal (ideally zero), in order to limit dephasing, while the transverse coupling $\beta_\perp$ remains
as large as possible, to optimize spin manipulation.
\\

\begin{figure*}[hbt!]
    \centering
    \includegraphics[scale = 1.15]{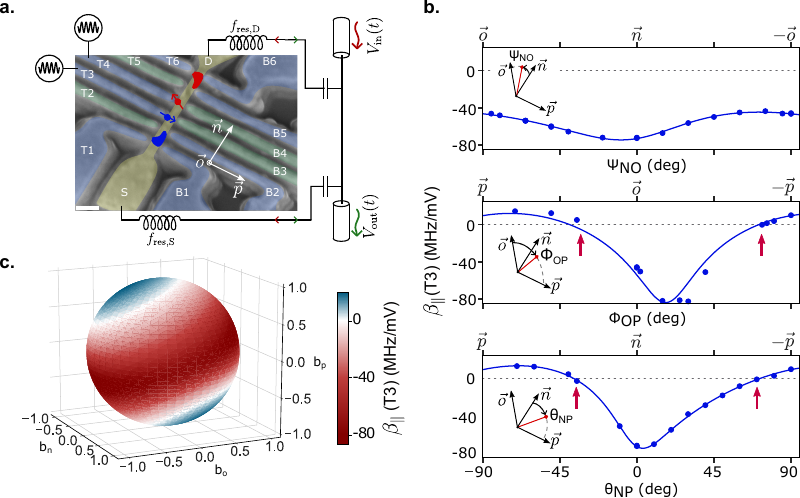}
    \caption{
    \textbf{Device and measurement of the longitudinal spin electric susceptibility (LSES)}
    \textbf{a} False-color scanning electron micrograph of the silicon metal–oxide–semiconductor device used in the experiment. It consists of a silicon nanowire channel (yellow) with p-doped source (S) and drain (D) contacts, and six gates on each side of the nanowire (T1,..,T6 and B1,..,B6).  The blue gates are negatively biased to accumulate holes, while the green gates are used to tune tunnel rates and confinement potentials. The investigated single-hole qubits, $Q_3$ and $Q_4$, are located next to the gates T3 and T4, respectively. $(\vec n, \vec o, \vec p)$ defines the coordinate system. 
    \textbf{b}  LSES of $Q_3$ to gate T3 denoted \lsesTthree, as a function of magnetic-field orientation in the three device symmetry planes. The red arrows mark the sweetspot orientations where \lsesTthree$= 0$. The solid line is a fit based on the g-matrix formalism (see Supplementary Information).
    \textbf{c} Full angular dependence of \lsesTthree as derived from the fits in \textbf{b}. The LSES is represented in color scale on the sphere defined by the magnetic-field unit vector, $(b_n, b_o, b_p)=\vec B/|\bm{B}|$. The LSES vanishes all along the white lines circling around the poles (sweetlines).
    \label{fig:1}
    }
\end{figure*}

\section{One qubit sweetline} \label{sec:sl}

Our experiment is performed on a metal-oxide-semiconductor (MOS) device consisting of an undoped natural silicon nanowire with p-doped ends connected to source and drain contacts and six finger gates on each side of the nanowire (see Fig. \ref{fig:1}\textbf{a}). We begin by forming a single-hole quantum dot at the lateral side of the nanowire accumulate by gate T3. In the presence of a magnetic field the accumulated hole encodes a two-level spin qubit, which we label $Q_3$. The single-shot readout of this qubit is performed by means of a well-established technique \cite{Elzerman_Nat2004}  based on spin-selective tunneling. The tunneling occurs to a large hole island defined in the nanowire portion near gates T1, B1, and B2, which are biased to a strong accumulation mode.  Tunneling is detected in real time by probing the charge state of the island via source-coupled, radio-frequency reflectometry (see Methods for more details on the readout technique and setup).    

Because of its proximity, T3 is the gate with the strongest effect on $Q_3$. Its voltage, $V_{\text{T}3}$, controls the perpendicular confinement of the dot hence has a strong influence on the effective g-factor anisotropy, which results in a generally large longitudinal spin-electric susceptibility (LSES), defined as  $\beta_\parallel(\text{T}3)=\partial f_L/\partial V_{\text{T}3}$. 
For this reason, and in line with previous observations on similar devices~\cite{Piot_2022}, we expect the hole spin qubit to be primarily sensitive to the perpendicular electric-field noise generated by charge fluctuators, probably located in the oxide of gate T3. To minimize the impact of this dominant noise contribution, we thus need to search for magnetic-field orientations where  $\beta_\parallel(\text{T}3)$ cancels out. 

Figure \ref{fig:1}\textbf{b} shows LSES measurements as a function of magnetic-field angle for three rotation planes (NO, OP, and NP) defined by the device symmetry axes, as indicated in Figure \ref{fig:1}\textbf{a}. Each data point is obtained by measuring the Larmor frequency at different gate voltages and fitting to a linear function (see Supplementary information). The LSES is clearly anisotropic and exhibit a pair of sweetspots ($\beta_\parallel(\text{T}3) = 0$) in two of the planes as indicated by the red  arrows.  The solid lines are fits based on the g-matrix formalism (Methods and \cite{Crippa_PRL_2018, Venitucci_2018}). From these fits we can reconstruct the full angular dependence of the LSES, which is displayed in Fig. \ref{fig:1}\textbf{c}. The LSES vanishes along two continuous lines circling around the axis of maximum $g$-factor (see Extended Data section for the full $g$-factor dependence). This result showcases the existence of a one-dimensional continuum of magnetic-field orientations where the qubit is resilient to the dominant electrical noise. The  observed ``sweetlines’’ are a generalization of the already reported sweetspots ~\cite{Piot_2022, Hendrickx_2024}. 

\section{Reciprocal sweetness}\label{sec:perf}

Usually, decoupling a qubit from the environment enhances its coherence time but lowers the qubit control efficiency with no net benefit for the qubit performance. This performance can be measured by the number of $\pi$ rotations within  the characteristic decay time $T_2^R$ of Rabi oscillations, the so-called gate quality factor, $Q$ \cite{Stano_2022}. The decoupling occurring on the experimentally observed sweetlines is however qualitatively different. As pointed out before, suppressing the LSES does not necessarily impact the ability to drive coherent Rabi oscillations. 
In fact, recent theoretical work \cite{michal_2023, Mauro_2023} has even shown that the transverse spin susceptibility maxima should lie on the sweetline, enabling an uncommon win-win scenario, referred to as ``reciprocal sweetness’’,  where the coherence time and the Rabi frequency are simultaneously enhanced, resulting in a larger quality factor.  

\begin{figure}[ht!]
	\centering
	\hfill\hbox to 0pt{\hss\includegraphics[width = \columnwidth]{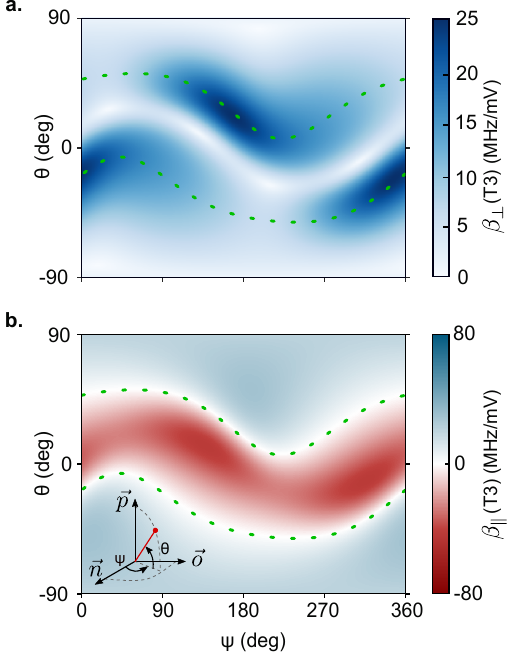}\hss}\hfill\null
	\caption{\textbf{Reciprocal sweetness}. 
Experimentally reconstructed relevant quantity for qubit $Q_3$, as function of magnetic field angle.
    \textbf{a} Color plot of the transverse spin electric susceptibility (TSES) of qubit $Q_3$ to gate T3, $\beta_\perp(\text{T}3)$, as a function of the magnetic-field angles $\theta$ and $\psi$, and for a constant qubit frequency $f_L = \SI{18}{\giga \hertz}$. The $\beta_\perp(\text{T}3)$ maxima (dark blue), which correspond to the highest Rabi frequencies, align with the sweetlines where \lsesTthree$=0$ (dashed lines). Conversely, the $\beta_\perp(\text{T}3)$  minima (light blue) coincide with the \lsesTthree maxima.
    \textbf{b} The angular dependence of \lsesTthree is reproduced here from Fig. 1 for direct comparison with $\beta_\perp(\text{T}3)$ in \textbf{a}. As in \textbf{a}, the sweetlines are indicated by dashed lines.}
    \label{fig:2}
\end{figure}

To experimentally test this prediction, we drive the qubit by applying a resonant voltage excitation to gate T3 and we determine the full angular dependence of the TSES, obtained from the ratio between measured Rabi frequency and voltage modulation amplitude. The results, generated from measurements and fits in the three symmetry planes (see Extended Data), are presented in Fig. \ref{fig:2}\textbf{a}, where $\beta_\perp(\text{T}3)$ is plotted in color scale as a function of the magnetic-field $\theta$ and $\psi$ angles (as defined in Fig. \ref{fig:2}\textbf{b}). The green dotted lines are the sweetlines obtained from the LSES measurements, which we plot onto Fig. \ref{fig:2}\textbf{b} to ease for the direct comparison with Fig. \ref{fig:2}\textbf{a}. The maxima of $\beta_\perp(\text{T}3)$ (dark blue) denote the locations of highest driving speed and they indeed lie on the sweetlines demonstrating the theoretically expected reciprocal sweetness.  By operating the qubit at these special locations, we leverage the two key properties of hole spins: their ability to hide from surrounding charge noise and their efficient response to electrical drive. As discussed in section \ref{sec:two_qubits}, this provides large quality factors. 

While there are only two absolute maxima of $\beta_\perp(\text{T}3)$ fully realizing the desirable reciprocal sweetness (one per sweetline), any vertical trace in Fig. \ref{fig:2}\textbf{a} exhibits at least a relative maximum, lying either on the upper or the lower sweetline.  Noteworthy,  $\beta_\perp(\text{T}3)$ displays a rather gentle dependence along a given sweetline, such that a fairly large maximum can be found for practically any azimuthal angle. 

\section{Single qubit tunability}\label{sec:tunability}

\begin{figure*}[hbt!]
	\centering
	\hfill\hbox to 0pt{\hss\includegraphics[width = \textwidth]{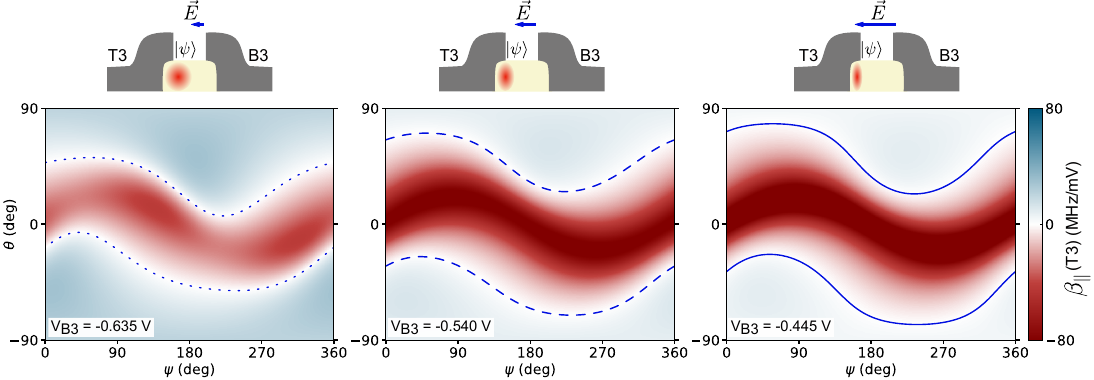}\hss}\hfill\null
	\caption{\textbf{Gate tunability of the sweetlines}.
Angular dependence of  \lsesTthree for three different gate-voltage settings at $f_L = \SI{17.99}{\giga \hertz}$). 
Increasing the voltage on gate B$3$ pushes the hole wave function against the left-side facet of the silicon nanowire as schematically represented in the top insets. The deformation of the wave function results in a shift of the sweetlines (blue lines). 
}
	\label{fig:3}
\end{figure*}

Figure \ref{fig:2} provides a ``user guide'' for the hole spin qubit, indicating how to tune it to its optimal operation point. In a hypothetical quantum processor, all the qubits are subject to the externally applied magnetic field. As a result, their performance can be simultaneously enhanced if they share a common optimal operating point. However, this condition is disrupted by charge disorder, which introduces variability in the hole confinement potential, in turn leading to deviations in the position of the qubit sweetlines. Here we use $Q_3$ to test the possibility to counteract such deviations by adjusting the gate voltages that control the qubit electrostatics. In fact, due to spin-orbit coupling, the g-factor anisotropy is expected to depend on the confinement potential of the quantum dot~\cite{Ares_2013, Voisin_2016, Liles_2021}. In the case of $Q_3$, confinement is mostly affected by the electric field orthogonal to the nanowire, which is controlled by the voltage difference between the facing gates T3 and B3. As a result, it can be shown that $\beta_\parallel(\text{T}3) \approx -\beta_\parallel(\text{B}3)$ (see Supplementary Information).

Figure \ref{fig:3} displays the LSES angular dependence measured for three different values of the voltage applied to gate B$3$ (meanwhile, $V_{\text{T}3}$ is simultaneously adjusted to keep the electrochemical potential of $Q_3$ dot approximately constant). 
Upon increasing $V_{\text{B}3}$, the hole wavefunction is squeezed more and more against the nanowire left-side facet. The effect is qualitatively illustrated in the upper insets, where the wave function deformation has been deliberately magnified for clarity. This deformation changes the anisotropy of the hole g-factor leading to a significant variation in the position and shape of the sweetlines. For a fixed azimuthal angle, we observe a shift of $10^\circ$ to $26^\circ$ in the polar coordinate of the corresponding sweetspot.  
The demonstrated electrostatic tuning offers a practical approach to compensate for moderate levels of qubit variability and realize noise-resilient operation across a multi-qubit processor.  Furthermore, dynamic local tuning can be leveraged to selectively enable and disable the longitudinal coupling of hole spin qubits to EDSR drives \cite{Froning_NNano_2021} or to microwave resonators \cite{Michal_PRB_2021, Bosco_PRL_2021}.

\section{Aligning two qubits sweetspots}\label{sec:two_qubits}

\begin{figure*}[h!]
    \includegraphics[width = \textwidth]{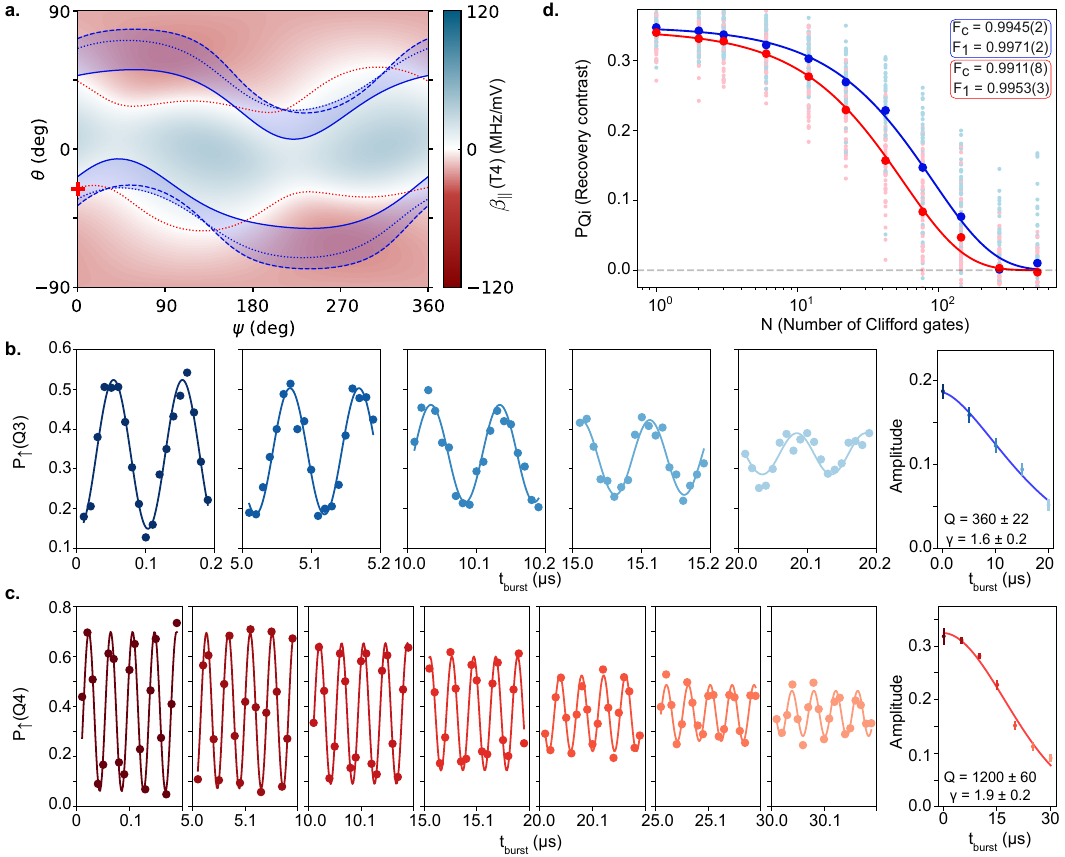}
	\caption{\textbf{Simultaneous tuning of two qubits to noise-resilient and fast operation points.}
  \textbf{a} LSES of qubit $Q_4$ to gate T4 as a function of magnetic-field orientation, with sweetlines indicated by dashed red lines. The previously measured sweetlines of qubit $Q_3$ (shown in Fig. 3) are reproduced here as dashed blue lines, with the shaded blue region in between highlighting the range over which qubit $Q_3$ can be tuned to a noise-resilient regime. 
  \textbf{b-c} Rabi oscillations measured at the common fast-operation sweetspot marked by the red cross in panel \textbf{a}. The different panels correspond  to widely spaced time intervals. Solid lines are fits to a sinusoidal function. The decay of the oscillation amplitudes are shown in the rightmost panels. Fitting to a generalized Gaussian decay function of the type $e^{-(t/T_R)^\gamma}$  (solid line) yields the Rabi decay time $T_R$ and hence the quality factor, $Q$, for single-qubit operation.  
 \textbf{d} Randomized benchmarking measurements for the two qubits, $Q_3$ (red) and $Q_4$ (blue), averaged over 40 repetitions shown as light red (blue) at the common sweetspot. $P_{Qi}$ is the recovery contrast of qubit $i$ measured after $N$ random Clifford gates (See Methods). $F_c$ is the fitted Clifford fidelity, and $F_1$ is the single qubit gate fidelity.
  }
  \label{fig:4}
\end{figure*}

We now present an experimental proof-of-concept demonstrating the feasibility of tuning two hole spin qubits to a shared optimal operation point. For this purpose, we form a second single-hole quantum dot, $Q_4$, confined by gate T4. Despite their proximity, $Q_3$ and $Q_4$ remain effectively independent due to negligible interaction. We begin by setting $V_{\text{B}3}=-\SI{0.58}{\volt}$, i.e. within the range explored in figure \ref{fig:3}, and tune  $Q_3$ to an operational sweetspot at $(\psi, \theta) = (0^\circ, 26^\circ)$, where the large transverse coupling term enables fast spin control.
Through a subsequent iterative procedure, we adjust the voltages on B$4$ and T4 in order to modify the electric field across $Q_4$ and hence achieve a vanishing LSES (i.e. \lsesTfour = 0). This straightforward procedure is sufficient to realize the desired sweetspot alignment. For completeness, we then perform an ensemble of additional LSES measurements at different magnetic-field orientations in order to reconstruct the full angular dependence of \lsesTfour, which is shown in Fig. \ref{fig:4}\textbf{a}. The red dotted lines highlight the sweetlines for $Q_4$. 
The blue lines are the three pairs of sweetlines belonging to $Q_3$ and previously shown in Fig. \ref{fig:3} for three different electrostatic configurations.
The chosen shared sweetspot at $(\psi, \theta) = (0^\circ, -26^\circ)$, indicated by a red cross, lies on the lower line. The shaded blue areas highlight the angular region over which $Q_3$ can be tuned to a noise-resilient regime, overlapping with the $Q_4$ sweetlines across  23 \% of the full $\psi$ range. This substantial overlap underscores the ample opportunity for simultaneous sweetspot operation.

We conclude this experimental study by investigating the performance of $Q_3$ and $Q_4$ at the common sweetspot through independent measurements of their Rabi oscillations.  
Both qubits exhibit a slow decay of the Rabi envelope with characteristic times $T_2^R(Q_3) = \SI{17.9(9)}{\micro \s}$ and $T_2^R(Q_4) = \SI{25.0(9)}{\micro \s}$, as shown in Fig. \ref{fig:4}\textbf{b-c}. They also operate at relatively large Rabi frequencies about $f_R(Q_3) \simeq \SI{10.0}{\mega \hertz}$ and $f_R(Q_4) \simeq \SI{24.0}{\mega \hertz}$, leading to remarkably high quality factors $Q = 2 \, T_2^R\,f_R$ of 360(20) and 1200(60) respectively. Noticeably, the quality factor of $Q_4$ surpasses the current state-of-the-art for semiconductor spin qubits~\cite{Yoneda_2018, Takeda_SciAdv_2016, Stano_2022}. To evaluate the single-qubit fidelities, we perform randomized benchmarking measurements whose results are presented in Fig. \ref{fig:4}\textbf{d}.  From the decay of the recovery contrast with increasing number of Clifford gates (N), we obtain averaged single-qubit gate fidelities of 99.53(3) and 99.71(2)\%  for $Q_3$ and $Q_4$ (see Methods). Both fidelities are above the fault-tolerant threshold for quantum error correction, yet, they are below the values that could be expected from the measured quality factors, which we respectively estimate to be about 99.85\% and 99.95\%. This discrepancy most likely results from a non-optimal calibration of the Clifford gates~\cite{Gustavsson_2013}. Further improvements in qubit performances could be achieved either by reducing the charge-noise level, currently an order of magnitude higher than those in state-of-the-art silicon-based devices or by replacing natural silicon with  $^{28}$Si-enriched silicon, in order to mitigate low-frequency magnetic noise from hyperfine interactions~\cite{Fischer_2008, Testelin_2009}. 

\section{CONCLUSIONS} 
The results of this work mark a pivotal advancement in understanding hole spin qubits and harnessing their favorable properties. The concepts of longitudinal and transverse spin-electric susceptibility, along with the observation of reciprocal sweetness, are expected to be broadly applicable to any type of spin-orbit qubit and semiconductor material.
Notably, this includes hole spin qubits in Ge/SiGe heterostructures, which have recently emerged as a promising platform, demonstrating remarkable progress and attracting increasing attention \cite{Wang_2024, Borsoi_2024, Jirovec_2021}.
A key takeaway from this study is that achieving optimal spin operation across an entire multi-qubit quantum processor appears feasible, provided the electrostatics of each qubit can be individually tuned and variability remains within manageable limits.

\bibliography{postdoc_bib.bib}

\section{Methods} \label{sec:methods}

\subsection{Device fabrication}
The device is a 6 split-gate silicon-on-insulator nanowire transistor fabricated in an industry standard 300 mm CMOS platform~\cite{Maurand2016}. The [110]-oriented silicon nanowire channel is $\SI{10}{\nano \meter}$ thick and $\SI{40}{\nano \meter}$ wide. It is connected on both ends to large boron-doped pads (labeled source and drain in the main text), used as reservoirs of positive charges. The 12 gates, etched on the top of the nanowire are $\SI{40}{\nano \meter}$ wide and spaced by $\SI{40}{\nano \meter}$ also. The gap between neighboring gates and between the doped contacts are filled with Si$_3$N$_4$ spacers. The gate stack consists in a $\SI{6}{\nano \meter}$ thick silicon oxide dielectric overlapped by a metallic bilayer of $\SI{6.5}{\nano \meter}$ of TiN and $\SI{50}{\nano \meter}$ of heavily doped poly-silicon.  A large metallic gate lying about $\SI{250}{\nano \meter}$ above the whole device is designed to globally shape the electrostatic landscape.

\subsection{Experimental setup}
The sample was cooled down at cryogenic temperatures 80-100 mK in a dilution refrigerator (Triton from Oxford Instrument) surrounded by a superconducting 3D vector magnet. DC voltages are provided by an IVVI DAC from Delft in the first 3 sections, then with BE2142 DACs from Bilt. Except for randomized benchmarking measurements, the microwave pulses are generated using a SMW200 vector signal generator (Rohde \& Schwarz), using frequency modulation (when measuring $\beta_\parallel$) or IQ modulation (for Rabi and Hahn-Echo experiments). Frequency (FM), IQ modulation as well as square pulses on T3 and T4 are generated using an Arbitrary Waveform Generator AWG5208 (Tektronix). A QBlox cluster was used for RB measurements, with a QCM module generating voltage pulses for gates T3 and T4, and a QCM-RF module used to generate the microwave signals. For readout, a high-frequency locking (UHF from Zurich Instruments) was used to generate and demodulate the 2 radio-frequency signals.

\subsection{Spin readout}
The spin readout of qubit $Q_3$ is performed with a secondary dot, located under gates T1, B1, and B2. At the readout conditions, the up-spin charge is energetically allowed to tunnel to a reservoir, which capacitively changes the electrical potential of the readout dot for a short time, until a new charge is reloaded on qubit $Q_3$. The readout dot is probed with a tank circuit connected to the nearby ohmic (S), itself driven at its resonance frequency $f_\text{REF, S} = \SI{361.5}{\mega \hertz}$. A decent reflectometry signal is achieved by tuning gates T1 and B1 for the readout dot to be coupled to the reservoir with a tunnel rate of few hundreds of MHz. Symmetrically, the qubit $Q_4$ is measured with the aid of a readout dot placed under gates T6, B6 and B5, probed at a frequency $f_\text{REF, D} = \SI{293.5}{\mega \hertz}$.

\subsection{g-matrix formalism}

We consider a single spin in a homogeneous magnetic field $\bm{B}=B{\bm b}$ oriented along the unit vector ${\bm b}$. The Hamiltonian of this spin can formally be written~\cite{Venitucci_2018}
\begin{equation}
\mathcal{H}_0=\frac{1}{2}\mu_B\bm{\sigma}\cdot\hat{g}\bm{B}=\frac{1}{2}g^*\mu_B B \sigma_\parallel\,,
                \label{eq:general_hamiltonian}
\end{equation}
with $\hat{g}$ a real $3\times 3$ matrix and $\bm{\sigma}\equiv(\sigma_x,\,\sigma_y,\,\sigma_z)$ the vector of Pauli matrices; $g^*=|\hat{g}\bm{b}|$ is the effective gyromagnetic factor and $\sigma_\parallel={\bm u}\cdot\bm{\sigma}$, where $\bm{u}=\hat{g}\bm{b}/g^*$ is the unit vector that defines the spin precession axis. We can moreover introduce the symmetric matrix $\hat{G}=\hat{g}^\mathrm{T}\hat{g}$ such that $g^*=\sqrt{\bm{b}\cdot\hat{G}\bm{b}}$; in hole spin systems, the spin precession (Larmor) frequency $f_L=g^*\mu_B B/h$ is usually anisotropic due to the strong spin-orbit coupling.

In a fluctuating gate voltage $\delta V$, this Hamiltonian can be further expanded as
\begin{equation}
\mathcal{H}(V_0 + \delta V) = \mathcal{H}_0 + h\delta V\left(\frac{1}{2}\beta_\parallel\sigma_\parallel+\beta_\perp\sigma_\perp\right)\,,
                \label{eq:ham_long_transv}
\end{equation}
where $\sigma_\perp={\bm v}\cdot\bm{\sigma}$ with ${\bm v}$ a unit vector orthogonal to ${\bm u}$. It features two distinct corrections: the first, $\propto\sigma_\parallel$ term is the longitudinal response, which describes how the Larmor frequency depends on the gate voltage but does not act on the spin precession axis. It is responsible for dephasing in noisy electric environments. The second, $\propto\sigma_\perp$ term is the transverse response, which tilts the spin precession axis and gives rise to Rabi oscillations (with Rabi frequency $f_R=\delta V\beta_\perp$ at resonance). Following Refs.~\cite{Venitucci_2018, michal_2023}, we can express the longitudinal ($\beta_\parallel$) and transverse ($\beta_\perp$) spin electric susceptibilities as a function of $\hat{g}$ (or $\hat{G}$) and their derivatives $\hat{g}^\prime=\partial\hat{g}/\partial V$ and $\hat{G}^\prime=\partial\hat{G}/\partial V$:
\begin{subequations}
\label{eq:ltses}
\begin{align}
    \beta_\parallel &= \frac{\partial f_L}{\partial V}
                    = \frac{\mu_B B}{2h g^*}\bm{b}\cdot \hat{G}^\prime\bm{b} \label{eq:lses} \\
    \beta_\perp &= \frac{f_R}{\delta V}=\frac{\mu_B B}{2h g^*}\left|(\hat{g}\bm{b})\times(\hat{g}^\prime \bm{b})\right|\,.
                \label{eq:tses}
\end{align}
\end{subequations}
Eqs.~\eqref{eq:general_hamiltonian} and \eqref{eq:ltses} were used to reconstruct the maps of $g^*$, LSES and Rabi frequencies of Figs. 1\textbf{c} and 2. For that purpose, the symmetric $\hat{G}$ and $\hat{G}^\prime$ matrices were fitted to the experimental Larmor frequencies and their dependence on gate T3 (Fig. 1\textbf{b}), while $\hat{g}^\prime$ was fitted to the experimental Rabi frequencies according to the methodology of Ref. \cite{Crippa_PRL_2018} (see supplementary materials).

\subsection{Randomized benchmarking}
Randomized benchmarking (RB) is used to assess the single-qubit gate fidelity \cite{Muhonen_2015,Lawrie_2023}. After initializing a spin-down state, we apply $N$ successive and randomly chosen gates from the Clifford group. This sequence is followed by a recovery gate, aiming at bringing the state alternatively to the up- or down-spin state. The contrast $P_{Q_i}$ is calculated from difference for 400 averages of 40 random choices. The contrast decays as $(2F-1)^N$ with $F$ the averaged Clifford gate fidelity. 
The Clifford set is constructed from physical gates, $\pi$ and $\pm\pi/2$ rotations around the $x$ and $y$ axes so as each Clifford gate concatenates on average 1.875 primitives. The single gate fidelity is therefore calculated as $F_1 = 1-(1-F_c)/1.875$~\cite{Muhonen_2015}.
The microwave pulses driving the qubit $Q_3$ (resp. $Q_4$) have a square envelope, and a duration of $\SI{48}{\nano \second}$ (resp. $\SI{20}{\nano \second}$) for the $\pi/2$ pulses and $\SI{96}{\nano \second}$ (resp. $\SI{40}{\nano \second}$) for the $\pi$-pulses.

\section{Extended data}

\begin{figure}[h!]
	\centering
	\hfill\hbox to 0pt{\hss\includegraphics[width = \columnwidth]{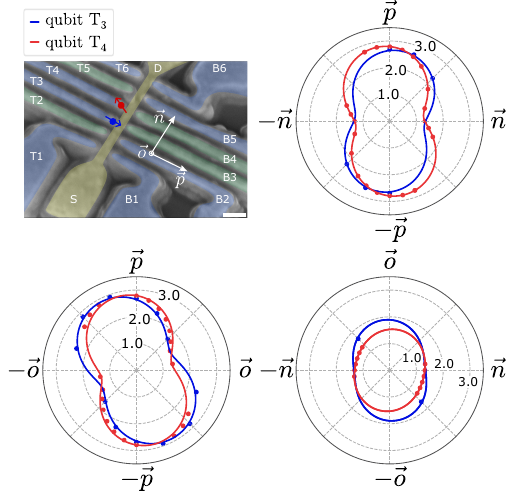}\hss}\hfill\null
	\caption[Hole effective g-factor comparison between qubits located below gates T3 and T4]{\textbf{Hole effective g-factor comparison between qubits located below gates T3 and T4:} 
  Blue (red) points are g-factor values evaluated by EDSR relating to the qubit located underneath gate T3 (T4). Solid lines are the fitted $g$-factor using the $g$-matrix formalism as presented in Methods. The $g$-factor configuration presented for the qubit $Q_3$ correspond to the dotted line configuration of  Figure 1.}
	\label{fig:g_matrices}
\end{figure}

\begin{figure}[h!]
	\centering
	\hfill\hbox to 0pt{\hss\includegraphics[width= \columnwidth]{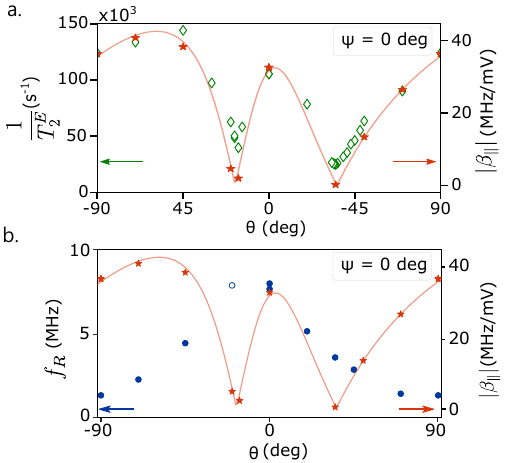}\hss}\hfill\null
	\caption[Hahn-Echo coherence time and Rabi frequency measured in the sample plane]{\textbf{Hahn-Echo coherence time and Rabi frequency measured in the sample plane:} 
  \textbf{a} Inverse of Hahn-Echo coherence time plotted as a function of magnetic field direction in the sample plane (NP). The red solid line represents the LSES fit (in absolute value) \lsesTthree as presented in the main text. 
  \textbf{b} Rabi frequency variations according to magnetic field direction in the sample plane (NP). Red line and points (fit) represent \lsesTthree in absolute value. At the sweetspot orientations two distinct cases are observed: one sweetspot with a fast electrical control ($\theta = \SI{-20}{\degree}$) and the second one ($\theta = \SI{35}{\degree}$) with a much slower Rabi frequency.}
	\label{fig:echo_np}
\end{figure}

\section*{Acknowledgments}
This work is supported by the French National Research Agency under the program ``France 2030'' (PEPR PRESQUILE - ANR-22-PETQ-0002), by the European Union's Horizon 2020 research innovation program (Grant Agreement No. 951852 QLSI) and the European Research Council (ERC) Project No. 810504 (QuCube). J.C.A.-U. is supported by Grants RYC2022-037527-I and PID2023-148257NA-I00 funded by MCIU/AEI/10.13039/501100011033 and by the ESF$+$. V.C. acknowledges support from the Program QuantForm-UGA  ANR-21-CMAQ-0003 France 2030 and by the LabEx LANEF  ANR-10-LABX-51-01.

\section*{Competing interests}
The authors declare no competing interests.

\section*{Data availability}
All data underlying this study are available in an online repository at XXX.

\section*{Additional Information}
Supplementary Information is available for this paper. Correspondence and requests should be sent to V. Schmitt (vivien.schmitt@cea.fr) and S. De Franceschi (silvano.defranceschi@cea.fr).

\end{document}